\documentstyle[12pt]{article}
\renewcommand{\theequation}{\arabic{section}.\arabic{equation}}
\renewcommand{\thesection}{ \setcounter{equation}{0} \Roman{section}.}

\renewcommand{\appendix}
 { \setcounter{section}{0}  \setcounter{subsection}{0}
 \renewcommand{\thesection}{ \setcounter{equation}{0}
 Appendix \Alph{section}:}
 \renewcommand{\theequation}{\Alph{section}.\arabic{equation}}}
\begin{document}
\begin{titlepage}
\begin{center}
{\large RUSSIAN ACADEMY OF SCIENCES \\ SIBERIAN BRANCH \\}
\vskip 1.5cm \
{\Large\bf The quartet state of the two-dimensional Heisenberg model with
the spin-$\frac{1}{2}$ on a square lattice.} \\
\vskip 1.cm
{\large V.I.Belinicher $^{*}$ and L.V.Popovich $^{+}$ }\\
\vskip 1.cm
$^{*}$Institute of Semiconductor Physics, 630090, Novosibirsk, Russia \\
$^{+}$Institute of Inorganic Chemistry, 630090,  Novosibirsk, Russia \\
(Received 1 June 1993)
\vskip 2.cm
\end{center}
\begin{abstract}
   The low-energy properties of the two-dimensional Heisenberg model with
spin-$\frac{1}{2}$ on a square lattice are investigated on the basis of the
local dimer order. The lattice is divided into square blocks consisting of
the quartet of spins. The spin variables and the Heisenberg Hamiltonian are
expressed in terms of the low-energy quartet variables. On the basis of the
Dyson-Maleev representation the spin-wave theory of the quartet state is
developed.  The spectrum of the lower magnon excitations consists of three
degenerate modes with the energy gap $\Delta =0.17J$. The ground state energy
per spin $E/N =-0.6J$. This preprint repeats in the main the previous one but
it contains calculations of the basic corrections and therefore has complete
character.
\vskip 0.5cm
PACS Numbers: 75.10.Jm, 75.30.Ds
\end{abstract}
\begin{center}
\vskip 2.cm
{\large Preprint N6}
\vfill

{\large Novosibirsk \\ 1993 }
\end{center}
\end{titlepage}

\newpage
 \section{Introduction} \ \
   Unusual states of the two-dimensional Heisenberg antiferromagnet attract
attention in connection with the problem of the magnetic state of the cuprate
superconductors \cite{1} and very complete references in this review.  This
activity was stimulated by the experiments which demonstrated that
superconductivity in cuprates is realized in the strongly correlated
paramagnetic spin state which was named the spin-liquid state.

Many different models of the spin-liquid state were proposed
\cite{1} - \cite{8}.  The
spin-liquid state as a linear combination of different dimer states was
considered in the works \cite{3,4} and was named the RVB-state.  The dimer
represents the state of the two spins-$\frac{1}{2}$ with the total spin equal
to zero.  The dimer can be formed from two neighboring spins-$\frac{1}{2}$ as
well as from two separated spins-$\frac{1}{2}$.  Numerical simulation of the
RVB-state was produced in the work \cite{4} and it was demonstrated that this
state for the two-dimensional Heisenberg antiferromagnet with the nearest
neighbor exchange had very low energy.  This energy is higher but very close
to the energy of the Neel state. Of course, the main problem is the
transformation of the magnetic state with doping.  But for progress in this
problem we must have sufficiently simple model of the spin-liquid state
because the original RVB-state permits only numerical consideration.

Such simple model of the spin-liquid state was proposed in the works
\cite{9,10}. The plane is split into square blocks containing four spins.
The block size is $2a$, where $a$ is the lattice constant.  The complete set
of eigenstates for such block or quartet consists of 16 states which can be
easily found for the exchange interaction $J_{1}$ between the first and
$J_{2}$ between the second neighbors.  For $J_{2}<J_{1}$ the ground state
$|\varphi >$  of such quartet is the state with the total momentum equal to
zero.  This state can be presented as a sum  of the two possible dimer states
for the nearest spins.  Its energy is $-2J_{1}+J_{2}/2 $.

As it was proposed
in \cite{9,10} that one can get good approximation for the spin-liquid state
if $|\varphi >$ state is chosen as ground state in each quartet and the
quantum fluctuations are taken into account.  At $ J_{2}=0 $ the energy per
site was in \cite{10} $ E_{0}=-0.655J $ that is very close to the numerical
results \cite{4} for  $E_{0}= -0.668J$.  In other paper \cite{9} based on the
quartet approach some sort of the Schwinger boson representation for the
Hubbard operators of a quartet was used.  In this work the energy was
sufficiently high $E_{0}=-0.57J$.  The correspondence between the numerical
simulation and the quartet model is not so obvious, because in the quartet
approach the quantum fluctuations are taken into account.  But approximations
of \cite{9,10} were sufficiently crude.  The
contribution of the higher-energy quartet states to the energy of the ground
state and to the spin dynamics was omitted.  In the work \cite{10}
uncontrollable approximation based on some decoupling scheme for the Green's
functions was used.  In the work \cite{9} the Schwinger boson method
is applicable only over parameter 1/N and it is also sufficiently crude.

In this work, which is based on the main idea of \cite{9,10} about the quartet
ground state for the Heisenberg antiferromagnet, we use spin-wave approach to
description of the triplet excitations. All approximations of this
work are controllable.  In this work we restrict our
consideration to the case of the unfrustrated Heisenberg antiferromagnet with
$J_{2}=0$. We consider contributions of the lower singlet state and all
triplet states into statistics and dynamics of the quartet state of the
Heisenberg antiferromagnet. The role of the additional singlet state and
the quintet state is essentially less because direct transitions from the
lower  singlet state into these states are absent. We do not consider these
states and it will be evident that their contribution into the ground state
energy is sufficiently less.

\section{The effective Hamiltonian of the model}\ \
The Hamiltonian of the Heisenberg antiferromagnet has a well-known form:
\begin{equation}
\label{1}
H=J\sum_{<l,l'>}({\bf s}_{l}{\bf s}_{l'})
\end{equation}
where $<ll'>$ notes summation over the nearest neighbors.  For construction
of the quartet magnetic state let us divide the square lattice into four-spins
blocks \cite{9,10}
\begin{equation}
\label{2}
\hat{Q}=\left(\begin{array}{cc} {\bf s}_{3} & {\bf s}_{2} \\
                                {\bf s}_{4} & {\bf s}_{1} \end{array} \right)
\end{equation}
and find all eigenstates of the quartet of spins (see Appendix A). Every
quartet has 16 states: two singlets $|\varphi >,\ |\psi >$ , three triplets
$|tm>$, $|xm>$, $|ym>$ and quintet $|qm>$. If we restrict our consideration
to the low-energy singlet $\varphi $ and triplet $t$ then the spins
${\bf s}_{i}$ consisting a quartet can be expressed in terms of the Hubbard
operators acting in $|\varphi>, |tm>,  m=\pm 1,0 $ subspace at the quartet
$n$:
\begin{eqnarray}
\label{3}
&&Z^{\varphi \varphi}_{n}=|\varphi n><n\varphi|,\ \ \ \ \
  Z^{tm,tm'}_{n}=|tmn><nm't|,
\nonumber\\
&&Z^{\varphi ,tm}_{n}=|\varphi n><nmt|,\ \ \
  Z^{tm, \varphi}_{n}=|tmn><n \varphi|,
\end{eqnarray}
The spin operator ${\bf s}_{ni}$ consisting of the quartet with the number
$n$ at the doubling square lattice may be represented as a sum of two vectors:
\begin{equation}
\label{4}
{\bf s}_{ni}=(-)^{i+1}{\bf L}_{tn}+\frac{1}{4}{\bf S}_{ttn}
\end{equation}
where ${\bf L}_{tn}$ mixes $|\varphi >$ and $|tm>$
states
\begin{eqnarray}
\label{5}
  L^{z}_{pn}=\frac{1}{\sqrt{6}}(Z^{\varphi,p0}_{n}+Z^{p0,\varphi}_{n}),\ \
  L^{+}_{pn}=\frac{1}{\sqrt{3}}(Z^{\varphi,p-1}_{n}-Z^{p1,\varphi}_{n}),\ \
  L^{-}_{pn}=\frac{1}{\sqrt{3}}(Z^{p-1,\varphi}_{n}-Z^{\varphi,p1}_{n})
\end{eqnarray}
for $p\ =\ t$ and ${\bf S}_{tt,n}$ are operators of the spin 1 acting in
t-subspace
\begin{eqnarray}
\label{6}
  S^{z}_{pqn}=Z^{p1,q1}_{n}-Z^{p-1,q-1}_{n},\ \
  S^{+}_{pqn}=\sqrt{2}(Z^{p1,q0}_{n}+Z^{p0,q-1}_{n}),\ \
  S^{-}_{pqn}=\sqrt{2}(Z^{p0,q1}_{n}+Z^{p-1,q0}_{n}),
\end{eqnarray}
for $p,q\ =\ t$, where $ A^{\pm} = A^{x}\pm iA^{y}$ are the spherical
components of a vector ${\bf A}$.  Using expression of (\ref{A4}) in Appendix
A for the energies of
$|\varphi >$ one can get the Hamiltonian of the problem
\begin{eqnarray}
\label{7}
&&H=-2J\sum_{n}Z^{\varphi ,\varphi }_{n}-J\sum_{nm}Z^{tm,tm}_{n} \nonumber \\
&&\ \ \ \ \ \ \
-(J/3)\sum_{<nn'>m}\left(2Z^{tm,\varphi }_{n}Z^{\varphi,tm}_{n'}+
  (-)^{m}(Z^{tm,\varphi}_{n}Z^{t-m,\varphi}_{n'}+Z^{\varphi,tm}_{n}
  Z^{\varphi,t-m}_{n'})\right) \nonumber \\
&&\ \ \ \ \ \ \
+(J/8)\sum_{<nn'>mm'}Z^{tm,tm'}_{n}\left(Z^{tm',tm}_{n'}-(-)^{m+m'}
Z^{t-m,t-m' }_{n'}\right).
\end{eqnarray}
The Hubbard operators $Z_{n}^{a,b}$ for $a,b =\varphi ,tm$ are not convenient
for the further analysis of the Hamiltonian (\ref{7}). Therefore we map the
Gilbert space consisting of $|\varphi >$ and $|tm>$ states into the Gilbert
space of the three-dimensional Heisenberg algebra $ t_{m}^{+}, t_{m}.$  Where
$t_{m}^{+}(t_{m})$ are creation (annihilation) Bose operators. The general
state of such algebra is $| n_{+}, n_{-},n_{0}>$ and mapping has simple form
\begin{eqnarray}
\label{8}
&&|\varphi>\rightarrow|0,0,0>,\ \ \ \ \ \ \ |t1>\rightarrow|1,0,0>,\nonumber \\
&&|t-1>\rightarrow|0,1,0>,\ \ \ |t0>\rightarrow|0,0,1>
\end{eqnarray}
This mapping generates the Holstein-Primakoff  representation \cite{11} for
our Hubbard operators
\begin{eqnarray}
\label{9}
&&(Z^{\varphi .\varphi }_{n})_{HP}=P(1-\hat{N}_{n})P,\ \ \ \ \ \ \ \ \
  (Z^{\varphi,tm}_{n})_{HP}=P\sqrt{1-\hat{N}_{n}}t_{nm}P, \nonumber \\
&&(Z^{tm,\varphi}_{n})_{HP}=Pt^{+}_{nm}\sqrt{1-\hat{N}_{n}}P,\ \ \
  (Z^{tm,tm'}_{n})_{HP}=Pt^{+}_{nm}t_{nm'}P, \nonumber \\
&&\hat{N}_{n}=\sum_{m}t^{+}_{nm}t_{nm},
\end{eqnarray}
where $P$ is a projector on the lower four states (\ref{8}) of the Heisenberg
algebra $t_{nm}^{+}, \ t_{nm}$. At the low temperatures $T \ll J$ or a small
excitation of oscillators $|nm>$ one can produce an isometric transformation
at each site
\begin{equation}
\label{10}
(Z^{a,b})_{DM}=V^{-1}(Z^{a,b})_{HP}V
\end{equation}
and omit the projector $P$. One can choose the operator $V$  from the
condition of elimination of roots in (\ref{10}), and get the Dyson-Maleev
representation for the Hubbard operators \cite{12,13}
\begin{eqnarray}
\label{11}
&&(Z^{\varphi ,\varphi }_{n})_{DM}=(1-\hat{N}_{n}),\ \ \ \ \ \ \ \ \
  (Z^{\varphi,tm}_{n})_{DM}=t_{nm}\nonumber, \\
&&(Z^{tm,\varphi}_{n})_{DM}=t^{+}_{nm}(1-\hat{N}_{n}), \ \ \ \
  (Z^{tm,tm'}_{n})_{DM}=t^{+}_{nm}t_{nm'}, \nonumber \\
&&\hat{N}_{n}=\sum_{m}t^{+}_{nm}t_{nm}.
\end{eqnarray}
The Dyson-Maleev representation for $Z$-operators is not a direct operator
identity but it can be used at low temperatures and a small level of
excitations.  Substituting the Dyson-Maleev representation (\ref{11}) in the
Hamiltonian (\ref{7}) we get the following effective Hamiltonian
\begin{eqnarray}
\label{12}
&&H=-(NJ/2)+J\sum_{n}\hat{N}_{n}-(J/3)\sum_{<nn'>m}
\Big(2t^{+}_{nm}(1-\hat{N}_{n})t_{n'm} \nonumber \\
&&\ \ \ \ \ \ \
+(-)^{m}\left(t^{+}_{nm}(1-\hat{N}_{n})t^{+}_{n'-m}(1-\hat{N}_{n'})+
t_{nm}t_{n'-m}\right)\Big) \nonumber \\
&&\ \ \ \ \ \ \
+(J/8)\sum_{<nn'>mm'}t^{+}_{nm}t_{nm'}\left(t^{+}_{n'm'}
t_{n'm}-(-)^{m+m'}t^{+}_{n'-m}t_{n'-m'}\right),
\end{eqnarray}
where $N$ is the number of spins in the plane.  This Hamiltonian is
nonhermitian due to properties of the Dyson-Maleev representation (\ref{11})
but if we use it for the computation of Green's functions any contradictions
do not appear.
\section{The approximate solution of the effective Hamiltonian}\ \
If we consider only the quadratic terms in the Hamiltonian (\ref{12}) we can
easily diagonalize them with the help of $u-v$ transformation (these quadratic
terms are Hermitian) and get the square of the magnon energy in the momentum
representation
\begin{eqnarray}
\label{13a}
&&E^{2}_{\bf k}=J^{2}(1-8\gamma_{\bf k}/3), \ \ \ \ \
\gamma_{\bf k}=(\cos(2k_{x}a)+\cos(2k_{y}a))/2,
\end{eqnarray}
where $a$ is the distance between spins, ${\bf k}$ is the quasimomentum.  One
can easily see that $E_{{\bf k}}^{2}$ is negative at $\gamma _{\bf k}>3/8$.
This means that our system is unstable in the quadratic approximation and the
higher terms in the Hamiltonian (\ref{12}) must be taken into consideration
for solving the problem of stability.  We will treat the Hamiltonian
(\ref{12}) by the mean-field method.  The same can be done by the Green's
function method. The results are identical.  We seek the solution which is
invariant over the rotation and consider the following normal and abnormal
averages:
\begin{eqnarray}
\label{13}
&&<t^{+}_{{\bf k}m}t_{{\bf k'}m'}>=\delta _{m,m'}\delta({\bf k}-{\bf k'})
N_{{\bf k}}, \nonumber \\
&&<t^{+}_{{\bf k}m}t^{+}_{{\bf k'}m'}>=(-)^{m}\delta _{m,-m'}
\delta({\bf k}+{\bf k'})F^{+}_{{\bf k}}, \nonumber  \\
&&<t_{{\bf k}m}t_{{\bf k'}m'}>=(-)^{m}\delta _{m,-m'}\delta({\bf k}+
{\bf k'})F^{-}_{{\bf k}}.
\end{eqnarray}
Our approach based on the Dyson-Maleev representation (\ref{11}) is correct if
the contribution of the higher states at every lattice sites is small. Because
of that the following conditions must be fulfilled
\begin{equation}
\label{14}
(\bar{N}_{p},\bar{F}^{\pm}_{p})\ = \
\sum_{\bf k}(\gamma_{\bf k})^{p}(N_{\bf k},F^{\pm}_{\bf k})\ \ll\ 1,
\end{equation}
where the sum over ${\bf k}$ notes the normalized integral over the Brillouin
zone. In this approximation we can omit the contribution of the members of the
six order and get the following mean-field Hamiltonian:
\begin{equation}
\label{15}
H_{mf}=E^{0}_{mf}+J\sum_{{\bf k}m}\left(\alpha_{{\bf k}}t^{+}_{{\bf k}m}
t_{{\bf k}m}+ (-)^{m}(\beta_{{\bf k}}t^{+}_{{\bf k}m}t^{+}_{{\bf -k}-m}+
\delta_{{\bf k}}t_{{\bf k}m}t_{{\bf -k}-m})/2\right),
\end{equation}
where $E_{mf}^{0}$ is the initial ground state energy in the mean-field
approximation:
\begin{eqnarray}
\label{16}
&&E^{0}_{mf}=(NJ/2)\Big(-1+(3\bar{N}_{0}-
4\bar{N}_{1}-2\bar{F}^{-}_{1}-2\bar{F}^{+}_{1})/4 \nonumber \\
&&\ \ \ \ \ \ \ \ \ \
-(3/8)\sum_{\bf k}(2\alpha_{\bf k}N_{\bf k}+
\beta_{\bf k}F^{+}_{\bf k}
+\gamma_{\bf k}F^{-}_{\bf k})\Big)
\end{eqnarray}
and the coefficients $\alpha _{\bf k}, \beta _{\bf k}, \delta _{\bf k}$ are
\begin{eqnarray}
\label{17}
&&\alpha_{\bf k} = 4(3/4+4(\bar{N}_{1}+
\bar{F}^{+}_{1})+(4\bar{N}_{0}+
\bar{F}^{+}_{0}-1+3\bar{N}_{1}/4)\gamma_{\bf k})/3,  \nonumber\\
&&\beta_{\bf k}=4(2(\bar{N}_{1}+\bar{F}^{-}_{1})+
(8\bar{N}_{0}-1
-3\bar{F}^{-}_{1}/4)\gamma_{{\bf k}})/3,  \nonumber\\
&&\delta_{\bf k}=4(2\bar{F}^{+}_{0}-1-
3\bar{F}^{+}_{1}/4)\gamma_{{\bf k}}/3.
\end{eqnarray}
The quadratic part of the Hamiltonian $H_{mf}$ (\ref{15}) was obtained from
$H$ (\ref{12}) by computing all possible paired averages in the fourth order
terms of $H$ (\ref{12}) .  The initial energy $E_{mf}^{0}$ is determined from
the condition of coincidence of the average values of the Hamiltonian
(\ref{12}) and (\ref{15}). This procedure corresponds to determination of the
one-particle energy by the Green's function method in the one-loop
approximation. The Hamiltonian (\ref{15}) can be diagonalized with the help
of the $u-v-w$ transformation (this procedure is equivalent of solving the
Dyson equation for normal and abnormal Green's functions):
\begin{eqnarray}
\label{18}
&&t_{{\bf k}m}=u_{{\bf k}}b_{{\bf k}m}+(-)^{m}
v_{{\bf k}}b^{+}_{{\bf -k}-m}, \nonumber\\
&&t^{+}_{{\bf k}m}=u_{{\bf k}}b^{+}_{{\bf k}m}+(-)^{m}
w_{{\bf k}}b_{{\bf -k}-m}, \nonumber\\
&&u^{2}_{{\bf k}}-v_{{\bf k}}w_{{\bf k}}=1.
\end{eqnarray}
After this transformation the Hamiltonian (\ref{15}) takes the form:
\begin{equation}
\label{19}
H_{mf}=E_{mf}+\sum_{{\bf k}m}E_{{\bf k}}b^{+}_{{\bf k}m}b_{{\bf k}m}
\end{equation}
with the ground state energy
\begin{eqnarray}
\label{20}
&&E_{mf}=E^{0}_{mf}+(3/4)N\sum_{{\bf k}}(E_{{\bf k}}-J\alpha_{{\bf k}})
\end{eqnarray}
and
\begin{eqnarray}
\label{20a}
&&E_{\bf k}=JR_{\bf k},\ \ \ \ \
  R_{\bf k}=\sqrt{\alpha^{2}_{\bf k}-\beta_{\bf k}\delta_{\bf k}}.
\end{eqnarray}

The coefficients $u_{{\bf k}},v_{{\bf k}},w_{{\bf k}}$ have form:
\begin{eqnarray}
\label{21}
&u_{{\bf k}}=\sqrt{(\alpha_{{\bf k}}+R_{{\bf k}})/2R_{{\bf k}}},\ \ \ \ \ \ \
&v_{\bf k}=z_{\bf k}\beta_{\bf k},\nonumber\\
&z_{{\bf k}}=-\sqrt{|\alpha_{\bf k}-R_{\bf k}|/2|\beta_{\bf k}
              \delta_{\bf k}|R_{\bf k}},\ \ \ \ \
&w_{{\bf k}}=z_{\bf k}\delta_{\bf k}.
\end{eqnarray}
So far as the coefficients $ \alpha _{\bf k},\beta _{\bf k},\delta_{\bf k}$
depend on the overages $\bar{N}_{p}, \bar{F}_{p}^{\pm}$ and consequently the
coefficients $ u_{\bf k},v_{\bf k}, w_{\bf k}$ also depend on these
parameters we can get the closed system of equations if we substitute Eqs.
(\ref{18}) for $t_{{\bf k}m}$, $t_{{\bf k}m}^{+}$ into Eqs. (\ref{13}),
(\ref{14}) which determine $\bar{N}_{p}, \bar{F}_{p}^{\pm}.$
\begin{eqnarray}
\label{22}
&&N_{\bf k}=(\alpha_{\bf k}/R_{\bf k})(n_{\bf k}+1/2)-1/2, \nonumber\\
&&F^{\pm}_{\bf k}=-[(\delta_{\bf k},\beta_{\bf k})/R_{\bf k}](n_{\bf k}+1/2),
\end{eqnarray}
where $n_{{\bf k}}$ is the Planck function: $n_{\bf k}=(\exp{(\beta
E_{k})} -1)^{-1}$ and   $\beta =1/T $ is the inverse temperature.
The system of equations of (\ref{22}) was solved numerically (see Appendix B
for details) and we get the averages $\bar{N}_{p}, \bar{F}_{p}^{\pm}$ as a
function of the temperature.  We also computed the free energy of the quartet
state from (\ref{19}) and determined the magnon energy $E_{\bf k}$
\begin{eqnarray}
\label{23a}
&&E_{{\bf k}}=\sqrt{A+2B\epsilon+C\epsilon^{2}}, \ \ \ \ \ \
\epsilon=\gamma_{\bf k} \nonumber\\
&&E_{{\bf k}}\approx\Delta(1+{\bf k}^{2}\xi^{2}),
\ \ \ \ \ \ \ \mbox{at} \ \ \ ka \ \ll\ 1 \nonumber \\
&&\Delta=\sqrt{A+2B+C}, \ \ \ \ \ \ \ \xi^{2}=-(B+2C)a^{2}/8\Delta^{2}
\end{eqnarray}
All these parameters are presented in Table 1. \\
\begin{center}
\begin{tabular}{c c c c c c}                         \hline    \hline
$\ \ \ T/J$  \ \ \  &\ \ \ \ \ $ F/NJ$ &\ \ \ \ \ $\Delta/J$\ \ \ \ \
&\ \ \ \ \  $A/J^{2}$\ \ \ \ \  &\ \ \ \ \  $B/J^{2}$\ \ \ \ \
&\ \ \ \ \  $C/J^{2}$\ \ \ \ \   \\ \hline
0     & -0.567 &  0.174   &   3.474   &  -1.595   &  -0.254    \\
0.1   & -0.567 &  0.205   &   3.484   &  -1.592   &  -0.258    \\
0.3   & -0.570 &  0.416   &   3.525   &  -1.562   &  -0.229    \\
0.5   & -0.585 &  0.704   &   3.452   &  -1.412   &  -0.133 \\ \hline \hline
\end{tabular}
\vskip 1.cm
\begin{tabular}{ c c c c c c c}                        \hline \hline
$\ \ \ \ \ T/J\ \ \ \ \ $ &\ \ \ \ $N_{0}\ \ \ \ $ &\ \ \ \ $N_{1}\ \ \ \ $
&\ \ \ \ $ F^{+}_{0}$\ \ \ \ &\ \ \ \ $F^{+}_{1}$\ \ \ \
&$\ \ \ \ F^{-}_{0}$\ \ \ \ &\ \ \ \ $ F^{-}_{1}$\ \ \ \ \\ \hline
0     & \ 0.045 &  0.025 &  0.087     &  0.137    &  0.035    &  0.083  \\
0.1   & \ 0.046 &  0.026 &  0.087     &  0.137    &  0.034    &  0.083  \\
0.3   & \ 0.050 &  0.029 &  0.086     &  0.136    &  0.028    &  0.076  \\
0.5   & \ 0.068 &  0.030 &  0.079     &  0.131    &  0.011    &  0.055  \\
 \hline  \hline
\end{tabular}
\end{center}
\vskip 0.5cm
TABLE 1. The result of calculation of the free energy $F$, the energy gap
$\Delta$, the parameters of the magnon energy $A,B,C$ and the normal and
abnormal averages $\bar{N}_{p}, \bar{F}_{p}^{\pm}$.\\

We can see that all normal and abnormal averages are small and do not exceed
0.14. This property of our solution justifies dropping of the six-order term
in the effective Hamiltonian (\ref{12}). The smallness of these averages
justifies also using of the Dyson-Maleev representation for the Hubbard
operators.  Really, one can check that the occupancy $W_{n_{+}n_{-}n_{0}}$ of
the state $|n_{+},n_{-},n_{0}> $ of the Heisenberg algebra $t_{m}^{+},\
t_{m}$ for our $u-v-w$ transformation (\ref{18}) is determined by the formula
\begin{equation}
\label{23}
W_{n_{+}n_{-}n_{0}}=(1+N_{0})^{-3}\left(\frac{N_{0}}{1+N_{0}}
\right)^{n_{+}+n_{-}+n_{0}}
\end{equation}
and the  total occupancy of the unphysical states $W_{unp}$ is equal to
\begin{equation}
\label{24}
W_{unp}\approx(6N^{2}_{0}+4N^{3}_{0}+N^{4}_{0})/(1+N_{0})^{4}=0.044.
\end{equation}
All this prove that we found correctly the rotationally invariant ground state
of the Hamiltonian (\ref{7}).

\section{Corrections to the mean-field energy of the ground state.}\ \
For checking accuracy of our solution we compute the second order correction
to the ground state energy in the mean-field approximation (\ref{20}). This
correction is presented by the Feynman diagram in Fig.1. \newpage
\setlength{\unitlength}{1mm}
\begin{picture}(0,0)
\put(40,0){\special{em:graph diagr.msp}}
\end{picture}
\vskip 4.cm
Fig.1 The Feynman diagram for the second order correction to the energy of
the ground state.

We restrict our consideration to the case of the zero temperature. The part
of the Hamiltonian (\ref{12}) essential for computation of the correction of
Fig.1 after u-v-w transformation has form
\begin{eqnarray}
\label{x1}
H_{int}=&&(J/12)\sum_{{\bf k}_{i},m,n}
\{ [16u_{{\bf k}_{1}}u_{{\bf k}_{2}}v_{{\bf k}_{3}}\gamma_{{\bf k}_{4}}
(u_{{\bf k}_{4}}+w_{{\bf k}_{4}})+ \nonumber\\
&&3\gamma_{{\bf k}_{1}+{\bf k}_{2}}
u_{{\bf k}_{1}}v_{{\bf k}_{2}}(u_{{\bf k}_{3}}v_{{\bf k}_{4}}-
u_{{\bf k}_{4}}v_{{\bf k}_{3}}](-)^{n+m}b^{+}_{m{\bf k}_{1}}
b^{+}_{n{\bf k}_{2}}b^{+}_{-n{\bf k}_{3}}b^{+}_{-m{\bf k}_{4}})+ \nonumber \\
&&(b^{+}_{m\bf k}\rightarrow b_{m\bf k},u_{\bf k}\rightarrow w_{\bf k},
v_{\bf k}\rightarrow u_{\bf k})\}.
\end{eqnarray}
Here summation ${\bf k}_{1},\ {\bf k}_{2},\ {\bf k}_{3},\ {\bf k}_{4}$
is produced over the surface ${\bf k}_{1}\ +\ {\bf k}_{2}\ +\ {\bf k}_{3}
\ +\ {\bf k}_{4}\ =\ 0$.

The first term in the square brackets in (\ref{x1}) follows from the $L-L$
interaction and the second term follows from the $S-S$ interaction. The last
term in the figured brackets is the conjugated to the first term. On the base
of the Hamiltonian one can easily compute the second order correction of Fig.1
to the ground state energy. We compute the six-dimensional integral over ${\bf
k_{1}},
{\bf k_{2}}, {\bf k_{3}}$  by the Monte-Carlo method and getting the
following result $\delta E=0.005 NJ$.

This correction to the ground state energy is sufficiently small that verifies
correctness of our mean-field solution of the Hamiltonian (\ref{12}). The
validity of the perturbation theory is based on the smallness of angle
integrals over ${\bf k}$ because another numerical small parameters are
absent in our theory. In fact, the small parameter is $1/z$ where $z$ is the
number of the neighboring spin to our quartet of spins.

\section{Influence of higher energy levels of the quartet on the ground state
         energy}\ \
It is clear that transitions at the higher energy levels of the quartet
reduce the ground state energy of the quartet state of the two-dimensional
Heisenberg antiferromagnets. As was noticed above the most essential
contributions originated from the higher triplet states $|xm>, |ym>$
(see (\ref{A5}). In order to consider this additional to $|\varphi >$ and
$|tm>$ states we must modify Eqs.(\ref{4})-(\ref{6}) for the spin operators
${\bf s}_{i}$ taking into account contribution of $|xm>, |ym>$ states. On the
base of the quartet wave function (\ref{A5}) we get the following
representation for the spin operators ${\bf s}_{i}$
\begin{eqnarray}
\label{a}
4{\bf s}_{i}=4(-)^{i+1}{\bf L}_{t}-2({\bf L}_{x}-(-)^{i}{\bf L}_{y})
         +{\bf S}_{tt}+{\bf S}_{xx}+{\bf S}_{yy} \nonumber \\
         +z_{i}({\bf S}_{tx}+{\bf S}_{xt}-(-)^{i}({\bf S}_{ty}+{\bf S}_{yt}))
             +(-)^{i}({\bf S}_{xy}+{\bf S}_{yx})
\end{eqnarray}
where $z_{i}=(1,1,-1,-1)$ for $i=1,2,3,4$. The operator of the triplet
excitations ${\bf L}_{p}$ and the spin-$1$ operator ${\bf S}_{pq}$ for
$p,q=t,x,y; \ m=0,\pm1 $ are expressed by the same manner as in Eqs.(\ref{5}),
(\ref{6}). For the further analysis we replace the Habburd projection
operators $Z^{\varphi \varphi }_{n}, Z^{\varphi, pm}_{n}, Z^{pm, \varphi}_{n},
Z^{pm,qm'}_{n}$ by the more convenient Bose operators $x^{+}_{nm},y^{+}_{nm},
t^{+}_{nm}\ (x_{nm},y_{nm},t_{nm})$ creating (annihilating) bosons of $x,y,t$
type with projection $m$ at site $n$. The corresponding  Dyson-Maleev
representation for the Hubbard operators is of the form (\ref{11}).
\begin{eqnarray}
\label{b}
&&(Z^{\varphi \varphi }_{n})_{DM}=(1-\hat{N}_{n}),\ \ \ \ \ \ \ \ \ \ \
  (Z^{\varphi, pm}_{n})_{DM}=p_{nm}\nonumber, \\
&&(Z^{pm, \varphi}_{n})_{DM}=p^{+}_{nm}(1-\hat{N}_{n}), \ \ \ \
  (Z^{pm,qm'}_{n})_{DM}=p^{+}_{nm}q_{nm'}, \nonumber \\
&&\hat{N}_{n}=\sum_{p,m}p^{+}_{nm}p_{nm}.
\end{eqnarray}
Hamiltonian (\ref{1}) can be represented as
\begin{eqnarray}
\label{c}
H = &&J\sum_{n}({\bf s}_{1n}+{\bf s}_{3n})({\bf s}_{2n}+{\bf s}_{4n})+
  J \sum_{<n,n'_{x}>}({\bf s}_{2n}{\bf s}_{3n'_{x}}+
                     {\bf s}_{1n}{\bf s}_{4n'_{x}})+ \nonumber \\
&& J\sum_{<n,n'_{y}>}({\bf s}_{2n}{\bf s}_{1n'_{y}}+
                     {\bf s}_{3n}{\bf s}_{4n'_{y}})
\end{eqnarray}
where $<n,n'_{x}> (<n,n'_{y}>)$ means summation over the nearest neighbor
of block $n$ in ${\vec x}$ (${\vec y}$) direction. Using the representations
(\ref{a}),(\ref{b}) for the spin operators ${\bf s}_{i}$ one can rewrite the
Hamiltonian (\ref{c}) in the momentum representation:
\begin{eqnarray}
\label{d}
&&H=H_{t}+H_{x}+H_{y}+H_{tx}+H_{ty}, \nonumber \\
&&H_{x}=J\sum_{{\bf k}m}\left((2 + \eta_{\bf k}/3)x^{+}_{{\bf k}m}
      x_{{\bf k}m}+ (-)^{m}\eta_{{\bf k}}(x^{+}_{{\bf k}m}
      x^{+}_{{\bf -k}-m}+x_{{\bf k}m}x_{{\bf -k}-m})/6\right), \nonumber \\
&&H_{y}=J\sum_{{\bf k}m}\left((2-\eta_{\bf k}/3)y^{+}_{{\bf k}m}
      y_{{\bf k}m}- (-)^{m}\eta_{{\bf k}}(y^{+}_{{\bf k}m}
      y^{+}_{{\bf -k}-m}+y_{{\bf k}m}y_{{\bf -k}-m})/6\right), \nonumber \\
&&H_{tx}=(iJ/3)\sum_{{\bf k}m}(-)^{m}sin(2k_{x}a)\left(t_{{\bf k}m}+
      (-)^{m}t^{+}_{{\bf -k}-m}\right)\left(x_{{\bf -k}-m}+
      (-)^{m}x^{+}_{{\bf k}m}\right), \nonumber \\
&&H_{ty}=-(iJ/3)\sum_{{\bf k}m}(-)^{m}sin(2k_{y}a)\left(t_{{\bf k}m}+
      (-)^{m}t^{+}_{{\bf -k}-m}\right)\left(y_{{\bf -k}-m}+
      (-)^{m}y^{+}_{{\bf k}m}\right).
\end{eqnarray}
where $\eta_{\bf k}=(cos(2k_{\bf x}a)-cos(2k_{\bf y}a))/2$ and
$H_{t}$ is the Hamiltonian (\ref{12}), which was discussed previously.
In $H_{x},H_{y},H_{tx},H_{ty}$ we excluded from our consideration all terms
more than the second order by creation and annihilation operators.
These terms lead
in the mean-field approximation to some renormalization of coefficients in
the quadratic form (\ref{d}) which does not effect essentially on results
obtained below.

The further plan is as follows. We take into consideration
the Hamiltonians $H_{x}$ and $H_{y}$ precisely and consider the Hamiltonians
$H_{tx}$ and $H_{ty}$ as perturbation. We will check that such approach is
suitable for the computation of corrections to the ground state energy.

The Hamiltonians $H_{x}$ and $H_{y}$ can be easily diagonalized with help of
the $u-v$ transformation (\ref{7}). The coefficients $\alpha_{\bf k},
\beta_{\bf k}$ and $\delta_{\bf k}$ in this case are
\begin{eqnarray}
\label{e}
\alpha_{\bf k}=J(2 \pm \eta_{\bf k}/3), \ \ \ \ \
\beta_{\bf k}=\delta_{\bf k} = \pm J\eta_{\bf k}/3,
\end{eqnarray}
where the upper sign "$+$" corresponds to $H_{x}$ and the lower sign "$-$"
$H_{y}$. As a result we get the energy for the $x,y$ quasiparticles in a form
\begin{eqnarray}
\label{f}
\varepsilon^{x,y}_{\bf k}=\sqrt{\alpha^{2}_{\bf k}-\beta^{2}_{\bf k}}
=2J\sqrt{1\ \pm \eta_{\bf k}/3}\ \simeq \
J(2 \pm \eta_{\bf k}/3 - \eta^{2}_{\bf k}/36).
\end{eqnarray}
The average number of quasiparticles $x,y$ in the ground state is
sufficiently small
\begin{equation}
\label{g}
<n_{x}>=<n_{y}>=\sum_{\bf k}v^{2}_{\bf k}\approx\frac{1}{144}
\sum_{\bf k}\eta^{2}_{\bf k}=\frac{1}{576}\ll N_{0}.
\end{equation}
Hence, the contribution of $H_{x}$ and $H_{y}$ in the ground state energy can
be estimated as $10^{-3}J$ per spin.

Notice, that dependence of the energies $\varepsilon^{x,y}_{\bf k}$ on
${\bf k}$ as it follows from Eq.(\ref{g}) is sufficiently weak. This leads to
the strong interaction inside the $x-y$ sector of the quartet Gilbert space.
As it follows from Eq.(\ref{a}) there are nonlinear terms in the total
Hamiltonian of the spin-spin interaction which lead to the strong interaction
inside of the $x-y$ sector. These terms in the Hamiltonians have a
characteristic form:
\begin{eqnarray}
\label{h}
H_{int}=(2J/\sqrt{3})\sum_{\bf k}\gamma _{\bf k}
({\bf L}_{t,\bf k}({\bf S}_{xy,\bf -k}+{\bf S}_{yx,\bf -k})
\end{eqnarray}
This means that, in fact, we do not know  the structure of excitations in the
$x-y$ sector. But for computation of corrections to the ground state energy we
can use the zero-dimensional approximation for the Green's function of the
$x,y$ quasiparticles \cite{Lv}. In this approximation the energy of $x$ and $y$
particles is simply $2J$ (\ref{f}) and their Green's function can be easily
found. One can show that in result of the interaction  (\ref{h}) a form of
the Green's function in the energetic space became the Gauss form instead the
Lorentz form as in the case of the weak interaction. But what is the most
important for us - a position of the center of peak in the energetic space of
the
imaginary part of the Green's function does not change at
$\epsilon =2J$. The corrections to the ground state energy due to the
Hamiltonian (\ref{h}) are equal to zero in this approximation, because a
contribution of the condensate the $x-y$ particles are small Eq.(\ref{g}).

The Hamiltonians $H_{tx}$ and $H_{ty}$ lead to an equal contributions to the
ground state energy, which can be calculated in the second order of the
perturbation theory:
\begin{eqnarray}
\label{i}
\delta E_{mf}&& = -(N/2)\sum_{{\bf q}l}
               \frac{<0|H_{tx}|1t_{{\bf q},l},1x_{{\bf -q},-l}>
               <1t_{{\bf q},l},1x_{{\bf -q},-l}|H_{tx}|0>}
               {\varepsilon^{x}_{\bf q}+E_{\bf q}} \nonumber \\
             &&=\ -(J^{2}N/6)\sum_{\bf q}sin^{2}(2q_{x}a)\frac
              {(u_{\bf q}+w_{\bf q})(u_{\bf q}+v_{\bf q})}
              {\varepsilon^{x}_{\bf q}+E_{\bf q}}\ =\ -0.024JN ,
\end{eqnarray}
where $u,v,w$ are defined in (\ref{21}) and $N$ is the number of spins in a
plane. We see that the interaction of
$x$ and $y$ particles with $t$ particle decreases the ground state energy by
0.024$J$ per spin. This correction will be reduced if we take into
consideration excluded from  Hamiltonian (\ref{d}) terms of third and forth
order over the creation and annihilation operators.

\section{Properties of the solution and discussion}\ \
The energy per spin of the quartet state at the temperature equal to zero
is sufficiently high: $E/N =-0.6J$. The coefficients $B$ and $C$ in the
magnon energy (\ref{23}) are negative and the energy has minimum at
${\bf k}=0$ and possesses a gap. This gap is equal to $0.174J$ at $T=0$.
Because our excitations have a gap, any long-range order is absent in the
quartet magnetic state. We can easily calculate the simultaneous spin
correlation function using representation (\ref{4}) for spin operators and
the representation (\ref{11}) for the Hubbard operators:
\begin{eqnarray}
\label{25}
K^{\alpha\alpha'}_{ni,n'i'}=<s^{\alpha}_{ni}s^{\alpha'}_{n'i'}>=
\delta _{\alpha \alpha '}K_{ii'}({\bf r}),\ \ \ \ \
{\bf r}={\bf r}_{n}-{\bf r}_{n'},
\end{eqnarray}
where $i, i^{'} = 1,2,3,4$ numerate spins inside quartet, $\alpha,\alpha '
= 1,2,3$  are vector indices.  According to (\ref{4}) we have
\begin{equation}
\label{26}
K_{ii'}({\bf r})=(-)^{i+i'}<L^{z}_{n}L^{z}_{n'}>+\frac{1}{16}<
S^{z}_{n}S^{z}_{n'}>.
\end{equation}
Substituting (\ref{5}), (\ref{6}) in (\ref{26}) and using (\ref{11})
one can get
\begin{eqnarray}
\label{27}
&&K_{ii'}({\bf r})=\frac{1}{6}(-)^{i+i'}\left(2(1-F^{+}_{0}-4N_{0})
N({\bf r})+(1-2F^{+}_{0})
F^{-}({\bf r})+(1-8N_{0})F^{+}({\bf r})\right)  \nonumber\\
&&\ \ \ \ \ \ \ \ \ \ \ \ \ +\frac{1}{8}\left(N^{2}({\bf r})-
F^{-}({\bf r})F^{+}({\bf r})\right),
\end{eqnarray}
were $N({\bf r}), F^{\pm}({\bf r})$ are the Fourier images of correlators
$N_{\bf k}, F_{\bf k}^{\pm}$:
\begin{equation}
\label{28}
\left(N({\bf r}),F^{\pm}({\bf r})\right)=
\sum_{{\bf k}}\exp(-i{\bf kr})(N_{{\bf k}},F^{\pm}_{{\bf k}}).
\end{equation}
When we derived (\ref{27})  we take into consideration the second and fourth
order over $t^{+},\ t$ terms. The contribution of the six-order terms is small
over $\bar{N}_{p}, \bar{F}_{p}^{\pm}$. The long-range order is absent in our
quartet state and asymptotic behavior is determined by the gap in the magnon
spectrum $\Delta$ and the effective magnon mass. It can be easily found from
(\ref{27}), (\ref{22}), (\ref{17}) and (\ref{23}) :
\begin{eqnarray}
\label{29}
K_{ij}(r)=D(-)^{i+j}(a/\xi)^{2}(2\pi r\xi)^{-1/2}\exp(-r/\xi),
\ \ \ \ \ r\gg\xi
\end{eqnarray}
The constant $D$ is a function ${\bar N}_{p},{\bar F}_{p}^{\pm}$ and the
correlation length $\xi$ is determined by the magnon energy.
Taking into account interaction of $x$ and $y$ particles with $t$ particle
does not lead to the efficient modification of the correlation function since
this interaction is strongly reduced at small ${\bf k}$ due to the structure
of contributions of the $x,y$ particles into the correlation function which
are proportional to $sin(2k_{x}a)$ and $sin(2k_{y}a)$. Therefore, a corrections
to $K_{ij}$ is of the order of $\delta K_{ij}\ \sim \ (a^{2}/\xi^{2})K_{ij}
\ll K_{ij}$. The behavior of the correlation function at the large distance
is similar to an antiferromagnet with short-range order \cite{1,7,8} and
contains the "stagger" factor $(-)^{i+j}$.

Our results essentially differ from
results obtained in works \cite{9,10} where the quartet state was proposed. We
considered only the two-dimensional unfrustrated Heisenberg antiferromagnets.
Our consideration is quantitative and
does not contain unjustified assumptions.  We did not find any gapless
excitations as in \cite{6,9,10} . Moreover our magnetic excitations have the
similar structure as in \cite{10}, but the energy is quite different. The
ground state energy in \cite{10} is sufficiently low $E_{0}/N =-0.655J$.

We connect these contradictions with the crude method of solution of the
Hamiltonian (\ref{7}).  The method of decoupling of Green's functions used in
\cite{10} is rather indefinite and result depends on the method of decoupling.
It is reasonable to suppose that low energy obtained in \cite{10} as well as
in the Schwinger boson method \cite{9} is an accidental result of
approximation.  The corrections to these approximations are not small and can
destroy this low energy.  We believe that application of the Schwinger boson
method to the Hamiltonian  (\ref{7}) \cite{9} has only qualitative character
as well.
\section{Conclusion}\ \
Our consideration of the quartet state of the two-dimensional
antiferromagnetic Heisenberg model was sufficiently consistent. Following
to \cite{9} and \cite{10} we have solved exactly the problem for four spins and
determined the initial ground state and structure of higher excitations. At
the  next step we produced mapping of our quartet state into the Bose system
with many degrees of freedom. At the low temperatures and level of excitations
this Bose system is reduced to the system with the nonhermitian polynomial
Hamiltonian. We constructed the ground state of this Hamiltonian in the mean-
field approximation and found a nontrivial solution of the self-consistent
equations for the normal and abnormal averages. We calculated the basic
corrections to the mean-field approximation and showed that they are small.
We took into consideration the most essential higher triplet excitations and
calculated their contribution to the ground state energy. We also found that
these contribution are small. We did not consider the higher singlet and
quintet excitations but their contribution to the ground state energy
definitely small because the direct transitions from the initial ground state
into these excited states are absent. We did not find an unstable excitations
in the quartet state.

AS a result we found that the ground state energy is sufficiently high
$E_{0}/N =-0.6J$ per spin. We believe that this result has a quantitative
character due to the effective small parameter of the perturbation theory.
It is in our case $1/z$ where $z$ is the number of spins neighbors to the given
quartet. In practice, the corrections are small due to smallness of the angle
integrals over the Brillouin zone.

Therefore, it is followed from our result, that the quartet state of the two-
dimensional antiferromagnetic Heisenberg model can not be in competition with
the Neel state as it was suggested in \cite{10}. Is the quartet short range
order in competition with the Neel short range order for the doped
antiferromagnet? It is an open question at present.
\vskip 1.0cm
\noindent
{\Large \bf Acknowledgments} \\ \vskip 0.1cm
We would like to thank S.A.Kivelson and S.V.Maleev for the stimulating
discussions, A.F.Barabanov and L.A.Maksimov for the helpful discussions and
sending reprints of their works on the quartet model. This work was supported
partly by the Council on Superconductivity of Russian Academy of Sciences,
Grant No. 90214 and by the scientific-technical program "High-Temperature
Superconductivity" as part of the state program "Universities as Center
for Fundamental Researches".
\vskip 1.0cm
\appendix
\noindent
\section{Structure of a single quartet}\ \
Every quartet consists of the four spins and has the Hamiltonian
$H_{q}$ :
\begin{eqnarray}
\label{A2}
H_{q}=J({\bf s}_{a}{\bf s}_{b})=(J/2)\left(({\bf s}_{a}+{\bf s}_{b})^{2}-
      {\bf s}^{2}_{a}-{\bf s}^{2}_{b}\right),
\end{eqnarray}
where  ${\bf s}_{a}= {\bf s}_{1} +{\bf s}_{3}$, $ {\bf s}_{b} =
{\bf s}_{2} +{\bf s}_{4} $. The states of a quartet can be numbered by the
four quantum numbers: $ S$ is the total spin of a quartet $ {\bf S} =
{\bf s}_{a}+{\bf s}_{b}$,  $s_{a}, s_{b}$ are the spins of $a$
and $b$ subsystems and $m$ is the projection of the total spin.  The
general state is $|s_{a}s_{b}Sm>$ and we have the following
states:
\begin{eqnarray}
\label{A3}
&&|\varphi>=|1,1,0,0>,\ \ \ \ \ \ |tm>=|1,1,1,m>, \ \ \ |\psi>=|0,0,0,0>,
\nonumber \\
&&|qm>=|1,1,2,m>,\ \ \ |am>=|1,0,1,m>,\ \ |bm>=|0,1,1,m>.
\end{eqnarray}
The energies of these states are following:
\begin{equation}
\label{A4}
E_{\varphi}=-2J,\ \ \ E_{t}=-J,\ \ \ E_{a}=E_{b}=E_{\psi}=0,\ \ \ E_{q}=J.\ \ \
\end{equation}
It is convenient for our purpose to use linear combinations
\begin{eqnarray}
\label{A4a}
|xm>=(|am>-|bm>)/\sqrt{2}, \ \ \ \ \ \ |ym>=(|am>+|bm>)/\sqrt{2}
\end{eqnarray}
instead the states $|am>$ and $|bm>$.
In this work we restrict our consideration by the $\varphi$ -singlet
and the $a,\ x,\ y$- triplet.  The wave functions of this states can be
presented in a form:
\begin{eqnarray}
\label{A5}
&&|\varphi>=\frac{1}{\sqrt{12}}\Bigg[
\left(\begin{array}{cc} + & + \\ - & - \end{array} \right)+
\left(\begin{array}{cc} - & - \\ + & + \end{array} \right)+
\left(\begin{array}{cc} - & + \\ - & + \end{array} \right)+
\left(\begin{array}{cc} + & - \\ + & - \end{array} \right)
\nonumber \\
&&\ \ \ \ \ \ \ \ \ \ \ \ -2\left(\begin{array}{cc} + & - \\ - & +
\end{array} \right)-
2\left(\begin{array}{cc} - & + \\ + & - \end{array} \right)\Bigg],
\nonumber \\
&&|t1>=\frac{1}{2}\Bigg[
       \left(\begin{array}{cc} + & + \\ - & + \end{array} \right)+
       \left(\begin{array}{cc} + & - \\ + & + \end{array} \right)-
       \left(\begin{array}{cc} + & + \\ + & - \end{array} \right)-
       \left(\begin{array}{cc} - & + \\ + & + \end{array} \right)\Bigg],
\nonumber \\
&&|t0>=\frac{1}{\sqrt{2}}\Bigg[
       \left(\begin{array}{cc} - & + \\ + & - \end{array} \right)+
       \left(\begin{array}{cc} + & - \\ - & + \end{array} \right)\Bigg],
\nonumber \\
&&|t-1>=\frac{1}{2}\Bigg[
        \left(\begin{array}{cc} - & - \\ - & + \end{array} \right)+
        \left(\begin{array}{cc} + & - \\ - & - \end{array} \right)-
        \left(\begin{array}{cc} - & + \\ - & - \end{array} \right)-
        \left(\begin{array}{cc} - & - \\ + & - \end{array} \right)\Bigg]
\nonumber \\
&&|x1>=\frac{1}{2}\Bigg[
       \left(\begin{array}{cc} + & + \\ - & + \end{array} \right)-
       \left(\begin{array}{cc} + & - \\ + & + \end{array} \right)+
       \left(\begin{array}{cc} + & + \\ + & - \end{array} \right)-
       \left(\begin{array}{cc} - & + \\ + & + \end{array} \right)\Bigg],
\nonumber \\
&&|x0>=\frac{1}{\sqrt{2}}\Bigg[
       \left(\begin{array}{cc} + & + \\ - & - \end{array} \right)-
       \left(\begin{array}{cc} - & - \\ + & + \end{array} \right)\Bigg],
\nonumber \\
&&|x-1>=-\frac{1}{2}\Bigg[
        \left(\begin{array}{cc} - & - \\ - & + \end{array} \right)-
        \left(\begin{array}{cc} + & - \\ - & - \end{array} \right)-
        \left(\begin{array}{cc} - & + \\ - & - \end{array} \right)+
        \left(\begin{array}{cc} - & - \\ + & - \end{array} \right)\Bigg]
\nonumber \\
&&|y1>=\frac{1}{2}\Bigg[
      -\left(\begin{array}{cc} + & + \\ - & + \end{array} \right)+
       \left(\begin{array}{cc} + & - \\ + & + \end{array} \right)+
       \left(\begin{array}{cc} + & + \\ + & - \end{array} \right)-
       \left(\begin{array}{cc} - & + \\ + & + \end{array} \right)\Bigg],
\nonumber \\
&&|y0>=\frac{1}{\sqrt{2}}\Bigg[
      -\left(\begin{array}{cc} - & + \\ - & + \end{array} \right)+
       \left(\begin{array}{cc} + & - \\ + & - \end{array} \right)\Bigg],
\nonumber \\
&&|y-1>=-\frac{1}{2}\Bigg[
        \left(\begin{array}{cc} - & - \\ - & + \end{array} \right)-
        \left(\begin{array}{cc} + & - \\ - & - \end{array} \right)+
        \left(\begin{array}{cc} - & + \\ - & - \end{array} \right)-
        \left(\begin{array}{cc} - & - \\ + & - \end{array} \right)\Bigg],
\end{eqnarray}
where signs $\pm$  note the direction of corresponding spins.  The
state $|\varphi> $ can be also presented as a sum of two possible
dimer configurations:
\begin{eqnarray}
\label{A6}
&&|\varphi>=(|d12>\otimes |d34> + |d14>\otimes |d23>)/\sqrt{3}, \nonumber \\
&&|dab>=(|a+>\otimes |b->-|a->\otimes |b+>)/\sqrt{2}.
\end{eqnarray}
This representation shows that our primary state sufficiently close
to the RVB-state.

\section{Structure of selfconsistent equations}\ \
When solving the system of equations (\ref{22}) we transform the
two-dimensional
over ${\bf k}$ integral into the one-dimensional over the energetic variable
$\epsilon $ one.  This is possible because the dependence over ${\bf k}$ is
not explicit but only via $\gamma _{{\bf k}}$ :
\begin{eqnarray}
\label{B1}
&&\sum_{\bf k}f(\gamma _{{\bf k}})=\int\limits_{-1}^{1}\rho(\epsilon)
f(\epsilon)d\epsilon,\ \ \ \
\rho(\epsilon)=\sum_{\bf k}\delta(\epsilon -\gamma_{\bf k}),\ \ \ \
\int\limits_{-1}^{1}\rho(\epsilon)d\epsilon\ =\ 1.
\end{eqnarray}
One can easily get for  $\rho (\epsilon  )$
\begin{eqnarray}
\label{B2}
&&\rho (\epsilon )=\frac{2}{\pi ^{2}}K(\xi'),\ \ \ \ \
\xi'=\sqrt{1-\xi ^{2}},
\end{eqnarray}
where $K(\xi')$ is a complete  elliptic integral of the first kind. For the
computation of integrals over $\epsilon $ we have used the Gauss type formula:
\begin{equation}
\label{B3}
\int\limits_{-1}^{1}\rho(\epsilon)f(\epsilon)d\epsilon=
\sum^{n}_{i=1}c_{i}f(\epsilon _{i}). \\
\end{equation}
Which is precise for only polynomial of the order $2n-1$. The values
$\epsilon _{i}$ are determined by zeros of the orthonormalized polynomials
$n$-th order on the segment (-1,1) with the weight $\rho (\epsilon )$ . The
coefficients $c_{i}$ also can be expressed in terms of these polynomials
\cite{14}. Substituting the expressions (\ref{17}) for $\alpha_{\bf k},
\beta_{\bf k}, \delta_{\bf k}$ in the terms $\bar{N}_{p}, \bar{F}_{p}$ into
Eqs. (\ref{22}) we get the following selfconsistent system of the equations
\begin{eqnarray}
\label{B4}
&&\bar{N}_{p}=(3/4+4\bar{N}_{1}+4\bar{F}^{+}_{1})I_{p}
+(4\bar{N}_{0}+\bar{F}^{+}_{0}-1 + 3\bar{N}_{1}/4)I_{p+1}-
\delta_{p0}/2, \nonumber \\
&&\bar{F}^{-}_{p}=-(2\bar{N}_{1}+2\bar{F}^{-}_{1})I_{p}
-(8\bar{N}_{0}-1-3\bar{F}^{-}_{1}/4)I_{p+1},\nonumber \\
&&\bar{F}^{+}_{p}=-(2\bar{F}^{+}_{0}-1-3\bar{F}^{+}_{1}/4)I_{p+1}.
\end{eqnarray}
Here the integrals $I_{p}$ are determined by the equations
\begin{eqnarray}
\label{B5}
f_{p}\ =\ I_{p}-\sum_{\bf k}(\gamma_{\bf k})^{p}R^{-1}_{{\bf k}}
(n_{{\bf k}}+1/2)=0, \ \ \ \ \ p=0,1,2.
\end{eqnarray}
As we can see from (B4) the five unknown variables $\bar{N}_{0},\bar{N}_{1}
,\bar{F}_{0}^{+}, \bar{F}_{1}^{+}, \bar{F}_{1}^{-}$ can be expressed in terms
of the three integrals $I_{p}$ for $ p=0,1,2$ which are determined by
Eqs.(B5).  These variables $I_{p}$ were found by searching zeroes of the
function
\begin{equation}
\label{B6}
\Phi=(f_{0})^{2}+(f_{1})^{2}+(f_{2})^{2}.
\end{equation}
This method of solving of Eqs.(B4) is more effective than the direct solution
of these equations.

\end{document}